\documentclass[ floatfix]{revtex4}
\usepackage{amsmath}
\usepackage{color}
\usepackage{amssymb}
\usepackage{graphicx}
\usepackage[utf8]{inputenc}
\usepackage[english, russian]{babel}
\usepackage{textcomp}
\usepackage{color}

\begin{document}

\title{Reentrant superconductivity in UTe$_2$ }

\author{V.P.Mineev$^{1,2}$}
\affiliation{$^1$Univ. Grenoble Alpes, CEA, INAC, PHELIQS, GT, F-38000 Grenoble, France\\
$^2$Landau Institute for Theoretical Physics, 142432 Chernogolovka, Russia}

\begin{abstract}

The
reentrant superconductivity  is the peculiar phenomenon observed in  paramagnetic metal UTe$_2$ in magnetic field parallel to the hard magnetisation axis. It  is difficult to explain it 
in terms of  field dependent  intensity of magnetic fluctuations  like it is done for explanation of  the formally similar phenomena in the 
ferromagnetic uranium superconductors URhGe, UCoGe.
Extremely large initial slope of the upper critical field temperature dependence
suggests that the phenomenon has  quasi-two-dimensional nature. Indeed,  according to the recent  band structure calculations 
by Yuanji Xu et al, Phys.Rev. Lett. {\bf 123}, 217002 (2019)
the Fermi surface of 
 UTe$_2$ consists of two types slightly corrugated cylinders.
 The theory of reentrant superconductivity in UTe$_2$ based on its quasi-two-dimensional structure is presented here.  
 \end{abstract}

\date{\today}
\maketitle

\section{Introduction}
The superconductivity in the metallic compound  UTe$_2$ with orthorhombic structure has been discovered in  December 2018 by Nicholas Butch and colleagues
\cite{Ran-Science}.
Almost immediately the discovery has been confirmed by the mixed French-Japanese team \cite{Aoki2019}. Since that time dozens studies of different properties of this material have been published. The most impressive observation  \cite{Ran-NatPhys,Knebel2019} was that the superconducting state 
 UTe$_2$ in the magnetic field aligned along the $b$-axis, which is  perpendicular to
the direction of easy magnetisation parallel to $a$-axis, persists up to 34.5 T where the superconductivity is destroyed by the metamagnetic transition. 
The magnetic field first reduces transition temperature to superconducting state, then  in the field interval (10 T, 27 T) $T_c(H)$ is almost field independent
and finally near the metamagnetic transition the transition temperature increases practically recreating its zero field magnitude \cite{Pourret2020}, 
see Fig.1. 
 Even more astonishing was that sweeping magnetic fields above the metamagnetic transition through the angular range 20\textdegree{}- 40\textdegree{} from the $b$-axis towards the 
 $c$-axis reveals a superconducting phase inside the dome, with the maximum value exceeding 65 T, the maximum field possible in the measurements (see Fig. 1 a,c in Ref.3.) The critical temperature of transition in superconducting state
in this material is 1.5 K.  The extremely high upper critical field  at such modest critical temperature indicates that in this compound the paramagnetic depairing mechanism does not work and the superconducting state  is formed by the electron pairs with spin $S=1$. 
At the same time, the usual orbital suppression of superconductivity also occurs ineffective. 

The reentrant superconducting state
under magnetic field perpendicular to the direction of easy magnetisation  revealed in UTe$_2$ reminds phenomenon known  
in ferromagnetic uranium compounds  URhGe, UCoGe \cite{Flouquet2019}.
For instance, in UCoGe the magnetic field directed parallel to $b$-axis, which is perpendicular to the direction of spontaneous magnetisation along $c$-axis, 
first suppresses the critical temperature but then at higher fields the critical temperature begins to increase. On the contrary, the field directed along spontaneous magnetisation effectively suppresses the superconducting state.
The situation looks  as a magnetic field 
in direction perpendicular to the  direction of easy magnetisation   stimulates pairing strength and a  field directed along the easy magnetisation suppresses the pairing interaction. 
The observations are not compatible with ordinary electron-phonon mechanism formation of superconducting state.
This suggests that pairing interaction is produced by magnetic fluctuations that is the pairing amplitude is determined by the magnetic susceptibility. 
Magnetic moment saturates under magnetic field parallel to spontaneous magnetisation.  This leads to susceptibility suppression which yields  
the decrease of the electron effective mass. Both these effects produce the reduction of pairing interaction. 
As a result, in addition to the usual orbital unpairing mechanism, another mechanism is added that accelerates the suppression of superconductivity,
 and the upper critical field $H_{c2}(T)$ along the $c$-axis exhibits an  upward curvature \cite{Mineev2020}.
The magnetic field perpendicular to the direction of spontaneous magnetisation reduces the Curie temperature thereby increasing  susceptibility
in the direction of spontaneous magnetisation \cite{Mineev2017}. The enhancement of susceptibility causes the electron effective mass increase. Thus, a magnetic field perpendicular to spontaneous magnetisation strongly stimulates  the superconductivity providing the high field reentrance of superconducting state.

 UTe$_2$  is not a ferromagnet. Its low temperature magnetic susceptibility along $b$-axis keeps a constant value till the metamagnetic transition \cite{Miyake2019}.  
 The specific heat coefficient $\gamma=C/T$ proportional to electron effective mass is  also practically constant till the fields
 about 30 Tesla and its increase appears only near the metamagnetic transition \cite{Miyake2019}. Thus, the mechanism responsible for 
 the reentrance  of superconducting state  in uranium ferromagnets in huge magnetic fields perpendicular to spontaneous magnetisation
 is not available in UTe$_2$ and cannot serve as an explanation of  stability of superconductivity  in this compound.

 The Fermi surface of UTe$_2$  found in the 
 recently reported first-principles band structure calculations \cite{Xu2019}    consists of two separate electron and hole cylinders with axes parallel to $c$-direction. Thus, UTe$_2$ looks as quasi-two-dimensional metal with conducting layers parallel to $(a,b)$ plane.
  It is known \cite{Lebed'1998} that the magnetic field parallel to conducting layers of quasi-two-dimensional metal
  first suppresses the  superconducting state and then at magnetic  energy $\hbar\omega_c$ comparable with the hopping amplitude between the conducting planes starts to recreate  superconductivity. In high fields the electrons move almost free along the open trajectories perpendicular  to the conducting layers. Thus, the physical reason for the suppression  of the orbital mechanism of the Cooper pair destruction  is  the suppression of the modulation of electron motion caused by the crystal field, in other words, the suppression of the trajectories curvature.

  The observed experimentally \cite{Ran-Science,Aoki2019,Knebel2019,Pourret2020} abnormally large initial slope of the upper critical field temperature dependence (see Fig.1) serves as an other solid argument in support of quasi-two-dimensional nature of
phenomenon of reentrant superconductivity in UTe$_2$.  
   
  Here I apply the theory of superconductivity in quasi-two-dimensional metals developed in the papers  \cite{Lebed'1998,Mineev2000}  to description of the upper critical field temperature dependence in UTe$_2$.

 \section{Upper critical field }

 In the  magnetic field parallel to $b$-axis it is reasonable to choose  $(x,y,z)$ coordinate axes directed along $(c,a,b)$ crystallographic directions.
 The corresponding elementary cell of reciprocal space is limited by intervals $ -\frac{\pi}{d}<p_x<\frac{\pi}{d},~-\frac{\pi}{a}<p_y<\frac{\pi}{a},~-\frac{\pi}{b},~<p_z<\frac{\pi}{b}$.
In this coordinate system  according to the paper \cite{Xu2019} there are two different conducting bands with the Fermi surfaces  consisting  of two pairs  corrugated cylinders with axes parallel to $p_x$ direction
 located in the points $(0,0,\pm\frac{\pi}{b})$ and $(0, \pm\frac{\pi}{a},0)$. 
 Thus,  the conducting layers of quasi-two-dimensional metal  are parallel to the plane $(a,b)$ and located at a distance $ d$ from each other.
  We will consider simplified single band model taking into account only the first band 
 with following electron spectrum near the Fermi surface
 \begin{eqnarray}
 \xi({\bf p})=\frac{1}{2m}\left (p_y^2+\left (p_z\mp\frac{\pi}{b}\right)^2\right)-2t\cos( p_xd)-\varepsilon_F
, \end{eqnarray}
such that $ t\ll\varepsilon_F$.
In the magnetic field ${\bf H}=(0,0,H)$,  ${\bf A}=(-Hy,0,0)$ parallel to the $b$ direction 
the electron wave function has the form
\begin{equation}
\Psi(x,y,z)=\psi(p_x,y,p_z)\exp\left (ip_xx+i\left (p_z\mp\frac{\pi}{b}\right)z\right)
\end{equation}
and $\psi(p_x,y,p_z)$ obeys the Schr\"odinger equation
 \begin{equation}
\left [ \frac{1}{2m}\left(-\frac{d^2}{dy^2}+\left (p_z\mp\frac{\pi}{b}\right)^2\right)-2t\cos(p_xd-\frac{\omega_cy)}{v_F}\right ] \psi(p_x,y,p_z)= \varepsilon\psi(p_x,y,p_z),
 \end{equation}
 where  $\omega_c={ev_FdH}/{c}$, and $\hbar=1$.  
 
 We do not take into account the band splitting due to the Zeeman interaction. In case of equal spin triplet pairing the latter does not produce  paramagnetic suppression of  superconducting state.  However, yielding 
 the opposite shifts in the Fermi momenta of spin-up and spin-down bands it produces the corresponding shifts in the density of states and changes the critical temperature of transition to the superconducting state ( see for instance \cite{Mineev2020}). This effect is absent In quasi - 2D case:  the magnetic field  parallel to the conducting layers changes the Fermi momenta but does not change the density of states near the Fermi surface.

 The normal state electron Green function derived in the same manner as in the paper \cite{Lebed'1998} is
 \begin{eqnarray}
 G_{\tilde\omega_n}(\phi,p_x,y,y_1)=-\frac{im~ sgn~\omega_n}{p_{0y}}
  \exp\left [\mp\frac{m\tilde\omega_n}{p_{0y}}   \right] \exp[\pm ip_{0y}(y-y_1)]~~~~~~~~~~~~~~~~~~~~~~~~~~~
\nonumber\\
\times
\exp\left\{\pm\frac{ i\lambda p_0}{p_{0y}} 
 \sin\left[ \frac{\omega_c(y-y_1)}{2v_F}\right ]\cos\left [p_xd- \frac{\omega_c(y+y_1)}{2v_F} \right ] \right\}, ~~~~~~~~~\pm\omega_n(y-y_1)>0.
 \end{eqnarray}
 Here, 
 the Matsubara frequency $\omega_n=\pi T(2n+1)$ is shifted  $\tilde\omega_n=\omega_n +\frac{1}{2\tau} sgn~\omega_n$
  due to   attenuation of electronic states produced by scattering on impurities,
 interaction with magnetic fluctuations etc, $\lambda=\frac{4t}{\omega_c}$, $ p_{0y}=p_0|\sin\phi |$ and $p_0$ is the Fermi momentum.

The simplest  equal spin pairing state has the order parameter
\begin{eqnarray}  
 \Delta(\phi,y)=\psi(\phi)\eta(y).
  \end{eqnarray}
  Here,
 \begin{eqnarray}
\psi(\phi)=A\left(\cos\phi+\frac{\pi}{bp_0 }\right ),~~~~~~~~~\frac{\pi}{2}<\phi<\frac{3\pi}{2},\\
\psi(\phi)=A\left(cos\phi-\frac{\pi}{bp_0 }\right),~~~~~~~~~-\frac{\pi}{2}<\phi<\frac{\pi}{2},
 \end{eqnarray}
 where $A$ is the normalisation constant such that $\frac{1}{\pi}\int_{\pi/2}^{3\pi/2}\psi^2(\phi)d\phi=1$.
 This type of the order parameter belongs to either $B_{2u}$ or $B_{3u}$ irreducible representation of the orthorhombic point group \cite{Xu2019}.
  The treatment of the equal spin pairing state belonging to the $A_u$ and $B_{1u}$ representations is mathematically  more cumbersome.
  
  The function $\psi(\phi)$ is odd function in respect to the point $\frac{\pi}{2}$: $\psi(\frac{\pi}{2}+\phi)=- \psi(\frac{\pi}{2}-\phi)$, whereas the Green function is even one. Hence, the corresponding to the order parameter self energy is equal to zero \cite{Mineev2000}.

The linear equation for the function $\eta(y)$ determining the upper critical field or the critical temperature $T_c(H)$ of transition to the superconducting state has the form \cite{Lebed'1998,Mineev2000}
 \begin{eqnarray}
 \eta(y)=\tilde g\int_{\pi/2}^{3\pi/2} \psi^2(\phi)\frac{d\phi}{\pi} \int_{|y-y_1|>a|\sin\phi|} 
\frac{2\pi Tdy_1}{ v_F|\sin\phi|} \frac{\exp\left[-\frac{|y-y_1|}{|\sin\phi| l} \right]}
  {\sinh\left[ \frac{2\pi T|y-y_1|}{v_F|\sin\phi| }  \right ]}\nonumber\\
  {\cal I}_0\left\{\frac{2\lambda}{|\sin\phi|} \sin\left[ \frac{\omega_c(y-y_1)}{2v_F}\right ]
  \sin\left [\frac{\omega_c(y+y_1)}{2v_F} \right ]  \right\}\eta(y_1),
  \label{eq}
 \end{eqnarray}
 where ${\cal I}_0(x)$ is the Bessel function, $\tilde g=\frac{mg}{4\pi d}$ is the product of the density of states
and  the pairing amplitude $g$, $a$ is a small distance cut-off, and $l=v_F\tau$ is the mean free path. 
 
 \subsection{Critical temperature}
 
 In the absence of a magnetic field the equation
 \begin{equation}
 1=\tilde  g \int_{\frac{2\pi aT}{v_F}}^\infty\frac{dz}{\sinh z}\exp\left(-\frac{z}{2\pi T\tau}   \right )
 \end{equation}
  rewritten as 
 \begin{equation}
 \ln\frac{T_c}{T_{c0}}=\psi\left(\frac{1}{2}\right)-\psi\left(\frac{1}{2}+\frac{1}{4\pi T_c\tau}\right)
 \end{equation}
  determines  the critical temperature.
  Here, $\psi(x)$ is the digamma function,
 \begin{equation}
 T_{c0}=\frac{v_F}{\pi a} \exp\left (-\frac{1}{\tilde g}   \right )
\end{equation}
 is the critical temperature in a perfect crystal without impurities $l=\infty$. The superconducting state suppresses completely at $l<\frac{\gamma v_F}{\pi T_{c0}}$, where $\ln\gamma=C\approx0.577...$ is the Euler constant.

\subsection{Ginzburg-Landau region} 

Near the critical temperature $T\approx T_c\gg\frac{\omega_c}{2\pi}$ the  essential  region of integration  in Eq.(\ref{eq}) is limited  by inequality \\
$\delta y<\frac{v_F|\sin\phi|}{2\pi T}$. Out this region the sub-integral expression is exponentially small. Hence, the product $\frac{\omega_c}{v_F}\delta y<\frac{\omega_c}{2\pi T}|\sin\phi|\ll1$, and the argument of the Bessel function
\begin{equation}
\frac{2\lambda}{|\sin\phi|} \sin\left[ \frac{\omega_c(y-y_1)}{2v_F}\right ]
  \sin\left [\frac{\omega_c(y+y_1)}{2v_F} \right ] \approx \frac{2t}{\varepsilon_F}\frac{\delta y}{|\sin\phi|}\omega_c m\xi   <      \frac{2t}{\varepsilon_F}\frac{\omega_c}{2\pi T}v_Fm\xi\ll1,
  \end{equation}
  proves to be small.  Here, $\xi$ is the characteristic length  on which the function $\eta(y) $ changes.
If $\xi\gg\delta y$  then one can expand $\eta(y_1)\approx\eta(y)+\eta^\prime(y)(y-y_1)+\eta^{\prime\prime}(y)(y-y_1)^2/2$ under integral in Eq.(\ref{eq}),
and also expand the Bessel function $ {\cal I}_0(x)\approx1-x^2/4$. Thus, we come to the differential equation
\begin{equation}
 \left[\ln\frac{T_{c0}}{T}+\psi\left(\frac{1}{2}\right)-\psi\left(\frac{1}{2}+\frac{1}{4\pi T\tau}\right)\right ]\eta(y)=-\frac{C_\psi I(\alpha)}{2}\left (\frac{v_F}{2\pi T}\right)^2\eta^{\prime\prime}(y)+I(\alpha)\left(\frac{t\omega_cy}{\pi v_FT}  \right )^2\eta(y),
 \label{eq1}
\end{equation}
 where $\alpha=(2\pi T_c\tau)^{-1}$,
 \begin{equation}
 I(\alpha)=\int_0^\infty\frac{z^2dz}{\sinh z}\exp\left(-\frac{z}{2\pi T\tau}   \right )=4\sum_{n=0}^\infty\frac{1}{(2n+1+\alpha)^3},
 \end{equation}
 \begin{equation}
 C_\psi=\int_{\pi/2}^{3\pi/2} \psi^2(\phi)\sin^2\phi\frac{d\phi}{\pi}.
 \end{equation}
 The lowest eigen value of Eq.(\ref{eq1}) at $T\approx T_c$ is
 \begin{equation}
 \left[\ln\frac{T_{c0}}{T}+\psi\left(\frac{1}{2}\right)-\psi\left(\frac{1}{2}+\frac{1}{4\pi T\tau}\right)\right ]=\frac{\sqrt{C_\psi}I(\alpha_c)t\omega_c}{2\sqrt{2}\pi^2T_c^2}.
 \end{equation}
 In pure case $\alpha\approx\alpha_c=(2\pi T_c\tau)^{-1}$ we obtain
 \begin{equation}
\omega_{c2}(T)=\frac{ev_Fd}{c}H_{c2}(T)=\frac{4\sqrt{2}\pi^2}{7\zeta(3)\sqrt{C_\psi} t}\left ( T_{c0}-\frac{\pi\beta}{8\tau}  \right )(T_c-T),
\label{H}
 \end{equation}
 where $\zeta(x)$ is the Riemann zeta function, 
 \begin{equation}
 T_c=T_{c0}-\frac{\pi}{8\tau},
 \end{equation}
and 
 \begin{equation}
 \beta=2-\frac{90\zeta(4)}{7\pi^2\zeta(3)}\approx0.83~.
 \end{equation}
 
 \subsection{High field region}
 
 The linear temperature dependence of $H_{c2}(T)$ near $T_c$  changes to the  more fast increase at smaller temperatures. The formal continuation of the linear dependence Eq.(\ref{H}) to $T=0$  ( see Fig.1) yields
 \begin{equation}
 \omega^{lin}_{c2}(0)=\frac{ev_Fd}{c}H_{c2}^{lin}(0)\approx 10\frac{T_c^2}{t},
 \end{equation}
 which can be rewritten in dimensional units in the form
  \begin{equation}
 \frac{e\hbar}{m_0c}H_{c2}^{lin}(0)\approx \frac{10}{\hbar v_Fdm_0}\frac{T_c^2}{t}
 \end{equation}
convenient for numerical comparison with experiment.  Here, $m_0$ is the electron mass in vacuum.
According to the available experimental data
\cite{Ran-Science,Aoki2019,Knebel2019,Pourret2020}
the values $H_{c2}^{lin}(T=0)$ are from 25 to 30 Tesla,  $T_c=1.5$ K. This gives us the possibility to estimate the magnitude of the interlayer hopping integral
 \begin{equation}
 t\lesssim \frac{1}{\hbar v_Fdm_0}~(Kelvin).
 \end{equation}

 Using this estimation we see that  the combination $ \frac{8t}{\omega_c|\sin\phi|}$ in the argument of the Bessel function in Eq.(\ref{eq})  begins to be smaller than unity at fields
 \begin{equation}
H> H_0=\frac{8}{(\hbar v_Fdm_0)^2}~(Tesla)
 \end{equation}
 except the small interval of angles $(\pi-\frac{8t}{\omega_c}<\phi<\pi+\frac{8t}{\omega_c})$. 
 
 To estimate the field dependence of critical temperature at $H>H_0$ let us divide the interval of integration over  the angle $\phi$ in the Eq.(\ref{eq}) as
 follows
 \begin{equation}
\frac{1}{\pi} \int_{\pi/2}^{3\pi/2}(...)d\phi =\frac{2}{\pi} \int_{\pi/2}^{\pi}(...)d\phi =\frac{2}{\pi}\int_{\pi/2}^{\pi-\frac{8t}{\omega_c}}(...)d\phi +
\frac{2}{\pi}\int_{\pi-\frac{8t}{\omega_c}}^{\pi}(...)d\phi.
 \end{equation}

Thus, in the angles region $\frac{\pi}{2}<\phi<\pi-\frac{8t}{\omega_c}$ at fields $H> H_0$  one can decompose the Bessel function ${\cal I}(x)=1-x^2/4$ and substituting the fast oscillating trigonometric functions by
  their average values we obtain
  \begin{equation}
  {\cal I}_0\left\{\frac{2\lambda}{|\sin\phi|} \sin\left[ \frac{\omega_c(y-y_1)}{2v_F}\right ]  \sin\left [\frac{\omega_c(y+y_1)}{2v_F} \right ]\right \}
\approx 1-\frac{t^2}{(\sin\phi)^2\omega_c^2}.
  \end{equation}
 On the other hand,  in the angles interval  $\pi-\frac{8t}{\omega_c}<\phi<\pi$ it is enough to take in mind that  the Bessel  function is ${\cal I}(x)<1$.
 
 Making estimation of the integrals over angle $\phi$ we come to 
  the equation for the critical temperature 
\begin{equation}
 1=\tilde  g \left [1-{\cal O}\left(\frac{8t}{\omega_c}\right )\right]\int_{\frac{2\pi aT}{v_F}}^\infty\frac{dz}{\sinh z}\exp\left(-\frac{z}{2\pi T\tau}   \right )
\end{equation}
 or
 \begin{equation}
 \ln\frac{T_c}{T_{c0}(H)}=\psi\left(\frac{1}{2}\right)-\psi\left(\frac{1}{2}+\frac{1}{4\pi T_c\tau}\right),
 \end{equation}   
where
\begin{equation}
 T_{c0}(H)=T_{co} \exp\left (-\frac{1}{\tilde g}
  \frac{{\cal O}\left(\frac{8t}{\omega_c}\right )}{1-{\cal O}\left(\frac{8t}{\omega_c}\right )}   \right )_{H\gg H_0} 
 \longrightarrow~ T_{c0}.
\end{equation}
Hence, in pure case
\begin{equation}
T_c(H)=T_{c0}(H)\left(1-\frac{\pi}{8\tau T_{co}(H)}\right)_{H\gg H_0} 
 \longrightarrow~ T_{c}.
\end{equation}
Thus, in high enough fields the critical temperature of transition to superconducting state tends to its zero field value.

\section{Conclusion}

We have demonstrated that the quasi-two-dimensional model allows to describe the phenomenon of reentrant superconductivity in UTe$_2$ in magnetic fields parallel to the $b$-axis. The comparison of the linear upper critical field temperature dependence in the Ginzburg-Landau region with available experimental data gives  the  estimation  of  the hopping integral between the conducting layers. The smallness of its magnitude opens the possibility  to the superconducting state recreation in reasonably high magnetic fields. 

The treatment has several  simplifications. We have considered single band model with parabolic spectrum, whereas the Fermi surface found in the paper \cite{Xu2019} consists of sheets corresponding to one electron and one hole band with more complex spectrum. So, the given approach must be generalised  taking into account the real band structure. This probably will allow to explain  other peculiar observations 
made in UTe$_2$ such that  half  of conducting electrons  seemingly do not participate in the superconductivity \cite{Ran-Science} and already mentioned the existence of the reentrant superconductivity  in the extremely high magnetic fields 
 in  the angular range 20\textdegree{}- 40\textdegree{} from the $b$ axis towards the $c$ axis \cite{Ran-NatPhys}. 
 
 In our theory linear in respect of the order parameter  we worked with the superconducting state belonging to $B_{2u}$ or $B_{3u}$ representations of the orthorhombic group $D_{2h}$.
 In the recent preprint \cite{Kapitulnik2020} there was revealed the splitting of transition to the superconducting state on two subsequent transitions
 unnoticed in all earlier publications.  For  fields oriented along the $a$- and $b$-axes the splitting was found invariable in all fields. 
 There was suggested the following sequence of transitions \cite{Kapitulnik2020}. First, it is transition to the $B_{2u}$ state and then to the non-unitary  state
 with the order parameter presenting the sum of the order parameters belonging to $B_{2u}$  and $B_{3u}$ representations shifted in respect each other on the phase angle $\pi/2$. The  description of the second transition is out the applicability of linear 
 theory. However, the main conclusion about the stability of superconducting state in high fields parallel to $b$-axis is  valid also for non-unitary superconductivity with the order parameter presenting the combination  of the order parameters relating to  $B_{2u}$ and  $B_{3u}$ representations.
 
In conclusion I consider it my pleasant duty  to express my gratitudes to Jean-Pascal Brison for the useful discussions of results and to Sheng Ran  for his interest in the work.

\begin{figure}[p]
\includegraphics
[height=1.0\textheight]
{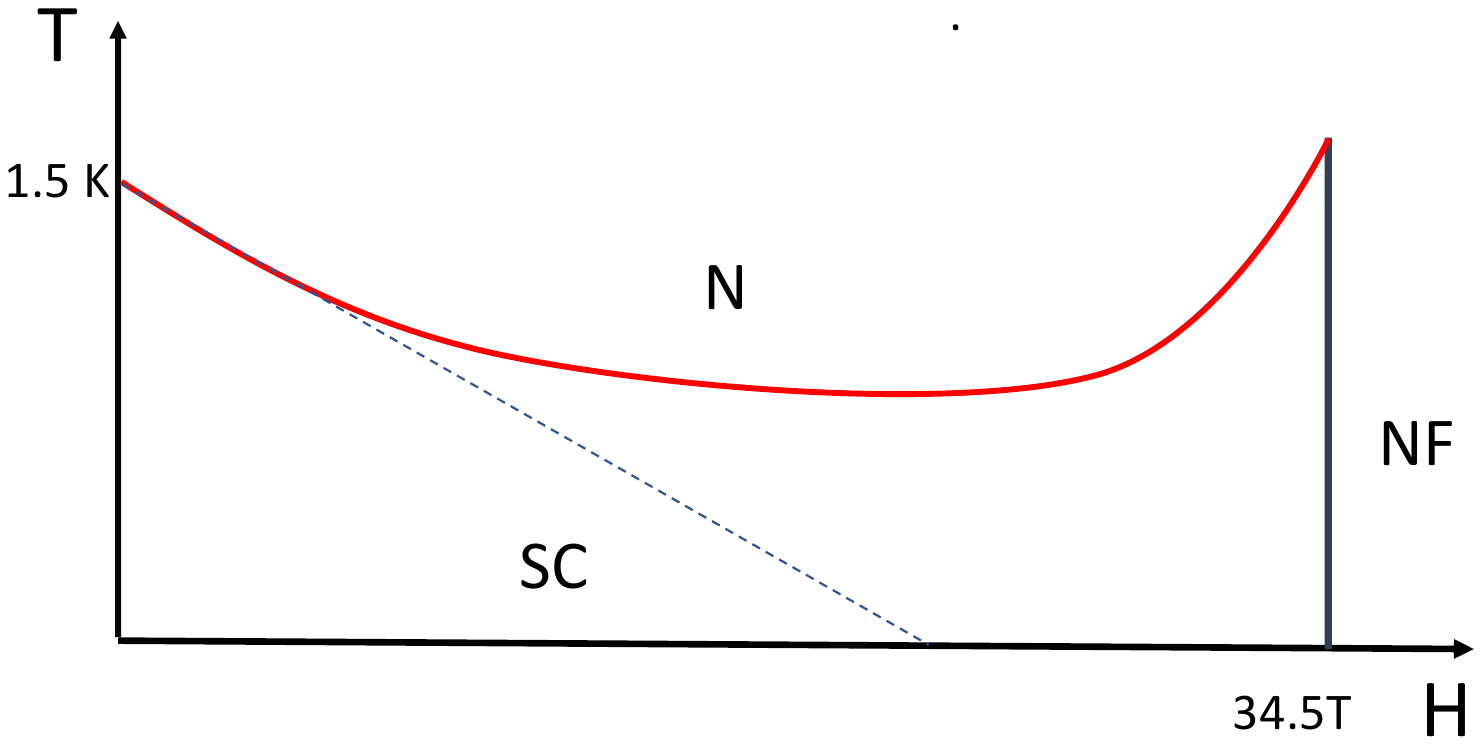}
 \caption{(Color online) 
Sketch of the critical temperature $T_c(H)$ magnetic  field dependence in UTe$_2$ for field parallel to $b$-axis
according to the paper \cite{Pourret2020}.
N, S, NF stand for  normal, superconducting and normal ferromagnetic phases correspondingly. Thin straight dashed line shows abnormally large initial slope of $H_{c2}(T)$ temperature dependence. }
\end{figure}


\begin{thebibliography}{220}

\bibitem{Ran-Science} S.Ran, C.Eckberg, Q.-P. Ding, Y.Furukawa, T.Metz, S.R.Saha, I-Lin Liu, V.Zic, H.Kim, J.Paglione, N.P.Butch, Science {\bf 365}, 684 (2019).

\bibitem{Aoki2019}D.Aoki, A.Nakamura, F.Honda, DeXin Li, Y.Homma, Y.Shimizu, Y.J.Sato, G.Knebel, J.-P.Brison, A. Pourret, D. Braithwaite, G.Lapertot, Qun Niu, M.Valiska, H.Harima, and J.Flouquet, J. Phys. Soc. Jpn. {\bf 88}, 043702 (2019).

\bibitem{Ran-NatPhys} S.Ran, I-Lin Liu, YunSuk Eo, D.J.Campbell, P.M.Neves, W.T.Fuhrman, S.R.Saha, C.Eckberg, H.Kim, D.Graf, F.Balakirev, J.Singleton, J.Paglione and N.Butch, Nature Physics {\bf 15},1250 (2019).

\bibitem{Knebel2019}G.Knebel, W.Knafo, A.Pourret, Qun Niu, M.Valiska, D.Braithwaite, G.Lapertot, M.Nardone, A.Zitouni, S.Mishra, I.Sheikin, G.Seyfarth, J.-P.Brison, D.Aoki, J.Flouquet, J. Phys. Soc. Jpn {\bf 88}, 063707 (2019).

\bibitem{Pourret2020}Q.Niu, G.Knebel, D.Braithwaite, D.Aoki, G.Lapertot, M.Valiska, G.Seyfarth, W.Knafo, T.Helm, J.-P.Brison, J.Flouquet, and A.Pourret, 
arXiv:2003.08986 [cond-mat] (2020).

\bibitem{Flouquet2019} D.Aoki, K.Ishida and J.Flouquet, J. Phys. Soc. Jpn. {\bf 88}, 022001 (2019).

\bibitem{Mineev2020} V.P.Mineev, Annals of Physics (NY), to be published (2020).

\bibitem{Mineev2017}V.P.Mineev, Usp. Fiz. Nauk {\bf 187}, 129 (2017) [Phys.-Usp. {\bf 60}, 121 (2017).

\bibitem{Miyake2019} A.Miyake, Y.Shimizu, Y.J.Sato, De Xin Li, A.Nakmura, Y.Homma, F.Honda, J.Flouquet, M.Tokunaga, and D.Aoki,
J. Phys. Soc. Jpn {\bf 88}, 063706 (2019).

\bibitem{Xu2019} Yuanji Xu, Yutao Sheng, and Yi-feng Yang, Phys.Rev. Lett. {\bf 123}, 217002 (2019).

\bibitem {Lebed'1998}A.G.Lebed' and K.Yamaji, Phys. Rev. Lett. {\bf 80}, 2697 (1998)

\bibitem{Mineev2000} V.P.Mineev, J.Phys.Soc.Jpn. {\bf 69}, 3371 (2000).

\bibitem{Kapitulnik2020}I.M. Hayes, Di S. Wei, T. Metz, Jian Zhang, Yun Suk Eo, S.Ran, S.R.Saha, J. Collini, N. P. Butch,  D.F. Agterberg, A. Kapitulnik, and J. Paglione, arXiv: 2002.02539.























\end{thebibliography}
\end{document}